\documentclass[referee]{raa}           

\usepackage{graphicx,times}
\usepackage{natbib}
\usepackage{amssymb,amsmath}
\bibpunct{(}{)}{;}{a}{}{,}

\usepackage[pagebackref=true]{hyperref}
\usepackage{bbding}

\usepackage{ulem}

\usepackage{color,xcolor}

\begin{document}
\title{Effect of Neutrinos on Angular Momentum of Dark Matter Halo}
\volnopage{ {\bf 20XX} Vol.\ {\bf X} No. {\bf XX}, 000--000}
\setcounter{page}{1}

\author{Yu Chen\inst{1,2,}\footnote[1]{The first two authors contribute equally to this work.}, 
Chang-Zhi Lu\inst{1,2,}$^\star$,
Yu Lu\inst{3},
Tingting Zhang\inst{4,}$^{\star\star}$, 
Tong-Jie Zhang\inst{1,2,5,}\footnote[2]{Corresponding author.},
}

\institute{Institute for Frontiers in Astronomy and Astrophysics, Beijing Normal University, Beijing 102206, China; {\it tjzhang@bnu.edu.cn}\\
\and
Department of Astronomy, Beijing Normal University, Beijing 100875, China\\
\and
Beijing Planetarium, Beijing Academy of Science and Technology, Beijing 100044, China\\
\and
College of Command and Control Engineering, PLA Army Engineering University, Nanjing 210017, China; {\it 101101964@seu.edu.cn}\\
\and 
Institute for Astronomical Science, Dezhou University, Dezhou 253023, China\\
\vs \no
{\small Received 20XX Month Day; accepted 20XX Month Day}
}

\abstract{Massive neutrinos are expected to affect the large-scale structure formation, including the major component of solid substances, dark matter halos. How halos are influenced by neutrinos is vital and interesting, and angular momentum (AM) as a significant feature provides a statistical perspective for this issue. Exploring halos from TianNu N-body cosmological simulation with the co-evolving neutrino particles, we obtain some concrete conclusions. First, by comparing the same halos with and without neutrinos, in contrast to the neutrino-free case, over 89.71\% of halos have smaller halo moduli, over 71.06\% have smaller particle-mass-reduced (PMR) AM moduli, and over 95.44\% change their orientations of less than $0.65^\circ$. Moreover, the relative variation of PMR modulus is more visible for low-mass halos. Second, to explore the PMR moduli of halos in dense or sparse areas, we divide the whole box into big cubes, and search for halos within a small spherical cell in a single cube. From the two-level divisions, we discover that in denser cubes, the variation of PMR moduli with massive neutrinos decreases more significantly. This distinction suggests that neutrinos exert heavier influence on halos' moduli in compact regions. With massive neutrinos, most halos (86.60\%) have lower masses than without neutrinos.
}
\authorrunning{Y. Chen \& C.-Z. Lu et al. }            
\titlerunning{Effect of Neutrinos on Dark Halo Angular Momentum}  
\maketitle

%
\section{Introduction}\label{sec1}
The angular momentum (AM) plays an important role in the formation of dark matter halos.
And dark matter halos in equilibrium exhibit some expected general properties, such as a universal density profile (the NFW profile, \cite{nava97apj}) of virialized halos in cosmological N-body simulations. The universal AM profile of dark matter halos was also carefully investigated by \cite{liao17apj} with Bolshoi cosmological simulation \citep{klyp11apj}. 
In such N-body simulation for dark matter halos, their measured AMs (although the plural form is Angular Momenta, we still use AMs for simplicity) will vary with the concrete particle sampling in mass resolution (or unit mass) configuration \citep{bull01mn,chen03apj}. Consequently, it is necessary to perform with a finer mass resolution for better computation on AM.

At present, neutrinos are the only detectable dark matter candidate for the usually extremely weak interplay with general substance \citep{liner17arx}, but their mass has not been known precisely.
A heavier neutrino can reduce the small-scale structure in the universe and change the power spectrum of total matter \citep{yu17natast}. Accordingly, neutrinos should only be a fraction of dark matter \citep{massey15mn}. Nowadays the lower limit on the sum of neutrinos' mass is $M_\mathrm{\nu}\equiv\sum_{\mathrm{i}=1}^{3}m_\mathrm{\nu_i}\gtrsim0.05$eV \citep{olive14chphc}, and the upper limit is $M_\mathrm{\nu}\lesssim0.12$eV \citep{planck20aap}.

The interaction between neutrinos and galactic dark matter halos via neutrino oscillation was explored by \cite{salas16prd}. Furthermore, \cite{yu19prd} believed that neutrinos make a unique contribution to the orientation of the AM fields of galaxies and halos. Based on the halos' AM profile research \citep{liao17apj}, it is quite natural to seek what effect neutrinos impose on the AMs of dark matter halos.

Massive neutrinos affect dark matter and large-scale structure in a linear perturbation \citep{lesgo06phr} at the early universe, but their influence now is highly non-linear \citep{zhu14prl} and complicated for analytic investigation \citep{casto15jcap,carbo16jcap}. Hence, our current knowledge of neutrinos dynamics mainly relies on numerical simulations. An N-body simulation with transparent mechanism, tremendous scale and high mass resolution \citep{liao17apj} of dark matter and especially neutrinos is entailed to our research. Although in some previous $N$-body simulations \citep{pears14phd}, neutrinos were set to have zero mass and contributed little to dark halos.

Fortunately, one of the world's largest N-body neutrino simulations, TianNu can be used for our research. Based on a flat $\Lambda$CDM model, including more than 3 trillion particles in a giant 1200$h^{-1}$Mpc-side-length cube, the TianNu simulation was run on Tianhe-2 supercomputer to obtain a precise CDM-neutrino dipole \citep{inman17prd} and to reveal the neutrinos' differential condensation according to their local dark halo density \citep{yu17natast}. Additionally, \cite{qin18apj} also used the TianNu data to research the effects of massive neutrinos on the CDM halos' spatial distribution in the form of Delaunay Triangulation (DT) voids. TianNu is an improved version of the public cosmological N-body code CUBEP$^3$M \citep{harno13mn} supporting the co-evolution of CDM and neutrinos.

The paper is organized as follows.
We make statistics of simulated halos in section \ref{sec2}, including their variation of AM moduli and orientations with and without neutrinos. We analyze the relation between halo AM modulus and its environmental density in section \ref{sec3}. Our discussions and conclusions are given in sections \ref{sec4} and \ref{sec5}.


\section{Methodology}\label{sec2}
\subsection{Data from TianNu Simulation}\label{sec2.1}
As for the TianNu simulation, all of its physical configurations are listed in \cite{yu17natast}, and all of its code details are available in \cite{ember17raa}. We just use TianNu and TianZero halos data and run the code of AM statistics.

Two independent N-body simulations, TianNu and TianZero, are performed for comparison. TianNu contains two light species in the background cosmology by CLASS transfer function \citep{blas11jcap} and a heavy one ($m_\mathrm{\nu}=0.05$eV) traced by simulated particles, while TianZero only discards massive neutrinos (or $m_\mathrm{\nu}=0$eV) with the same other conditions.

A flat $\Lambda$CDM is applied in the simulations. The density parameters of CDM ($\Omega_\mathrm{c}$) and baryons ($\Omega_\mathrm{b}$), reduced Hubble parameter ($h$), initial tilt ($n_\mathrm{s}$) and fluctuations of the power spectrum (mass variance at 8$h^{-1}$Mpc, $\sigma_\mathrm{8}$) are assigned as 0.27, 0.05, 0.67, 0.96 and 0.83 separately. Given neutrino density parameter $\Omega_\mathrm{\nu}$\footnote{Note here, $\Omega_\nu\equiv1.68h^{-2}\times10^{-5}$ for $\sum_\mathrm{i} m_{\nu\mathrm{i}}$=0, otherwise $\Omega_\nu\equiv(94h^2\mathrm{eV})^{-1}\sum_\mathrm{i} m_{\nu\mathrm{i}}$.}, the total matter density parameter is $\Omega_\mathrm{M}=\Omega_\mathrm{c}+\Omega_\mathrm{b}+\Omega_\mathrm{\nu}$, and it leads to $\Omega_\mathrm{\Lambda}=1-\Omega_\mathrm{M}$.

The simulation is conducted with only $6912^3$ CDM particle-groups (mass resolution, $\mathrm{M_{r,\nu}}=6.9\times10^8\mathrm{M_\odot}$ in TianNu and $\mathrm{M_{r,z}}=7\times10^8\mathrm{M_\odot}$ in TianZero) from z=100 to z=5.
At lower redshift, the TianNu is injected with $13824^3$ neutrino particle-groups (mass resolution $3\times10^5\mathrm{M_\odot}$) to co-evolve, while the TianZero keeps neutrino-free.
Under this preset of mass resolutions, the total masses of the two simulations are almost equal, making $\Omega_\nu\approx0.37\%\Omega_\mathrm{M}$$\approx0.0011$.
The simulation information is saved in 21 checkpoints including all positions and velocities, and the data we use are the checkpoints at z=0.01.

\begin{figure}[htb]
	\centering
	\includegraphics[scale=0.75]{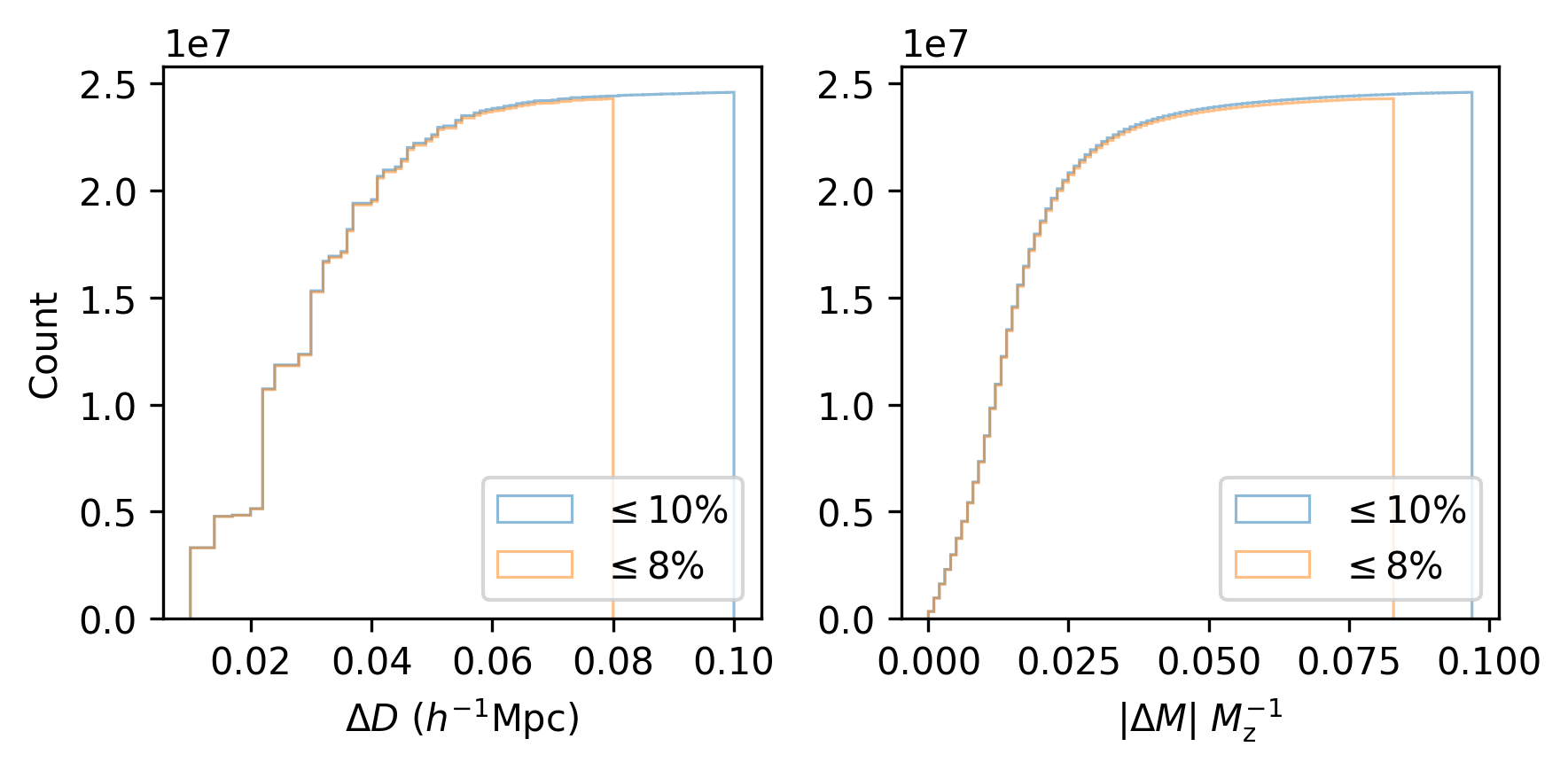}
	\caption{The cumulative count curves of distance- and percent mass difference between halo pairs in the left and right panel correspondingly. The $M$ is halo mass, and $\Delta M=\mathrm{abs}(M_\mathrm{\nu}-M_\mathrm{z})$. The blue curve is the 10\%-variation case, and the red curve is the 8\%-variation case taking up 97.6\% of the 10\% case.}\label{fig1}
\end{figure}

A Spherical Overdensity (SO) algorithm \citep{yu17natast} is used to discover halos with their mass in TianZero from about $9\times10^{11}$ to $3.6\times10^{14}\mathrm{M_{\odot}}$. The final data we get contain two halo sets (TianNu and TianZero), where the properties of each halo are described by seven parameters: its mass, three-dimensional orthogonal components of position and AM modulus.
The late-time injection of neutrinos makes little impact on the halo's mass and position, which can be seen in Figure \ref{fig1}. The 10\% variations of mass-center position and mass concerning TianZero halos were used in the former studies \citep{yu17natast,qin18apj}. However, considering the potential improvement of the program, we adopt an 8\% variation, summing up to a 94.9\% match rate and 24301359 pairs of halos.

\subsection{Calculation for AM}\label{sec2.2}
The AM $\bm{J}$ of a dark matter halo is obtained by summing up the AMs of all particles within it:
\begin{equation}\label{eq1}
	\bm{J} = \sum_{i=1}^N m_i \bm{r}_i \times \bm{v}_i  ,
\end{equation}
where $i$ represents a single particle with its mass $m_i$, coordinate $\bm{r}_i$, and velocity $\bm{v}_i$.
For the AM of every halo, we explore its modulus- and direction-difference individually between two datasets.

\begin{figure}[htb]
	\centering
	\includegraphics[scale=0.75]{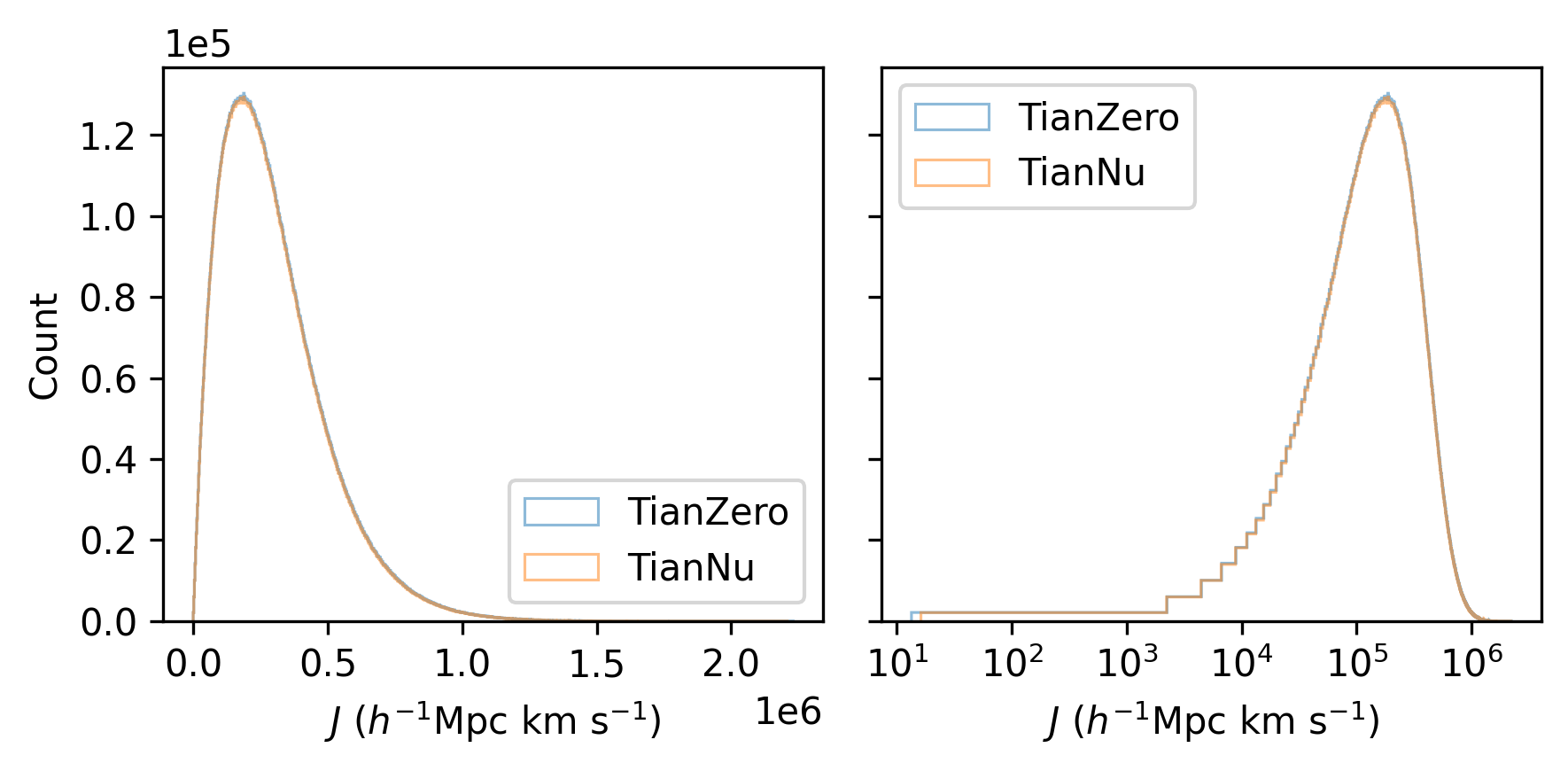}
	\caption{The count curves as functions of halo modulus. The blue and red curves are the TianZero and TianNu counts. The right panel plots the same distribution with logarithmic abscissa.}\label{fig2}
\end{figure}

We define the modulus of the summed AM from eq. \ref{eq1} as the halo modulus ($J=|\bm{J}|$), and plot them for all halos in Figure \ref{fig2}, showing highly similar distributions of the two sets. And a seemingly left-skewed lognormal function can fit the characteristics.

\begin{figure}[htb]
	\centering
	\includegraphics[scale=0.9]{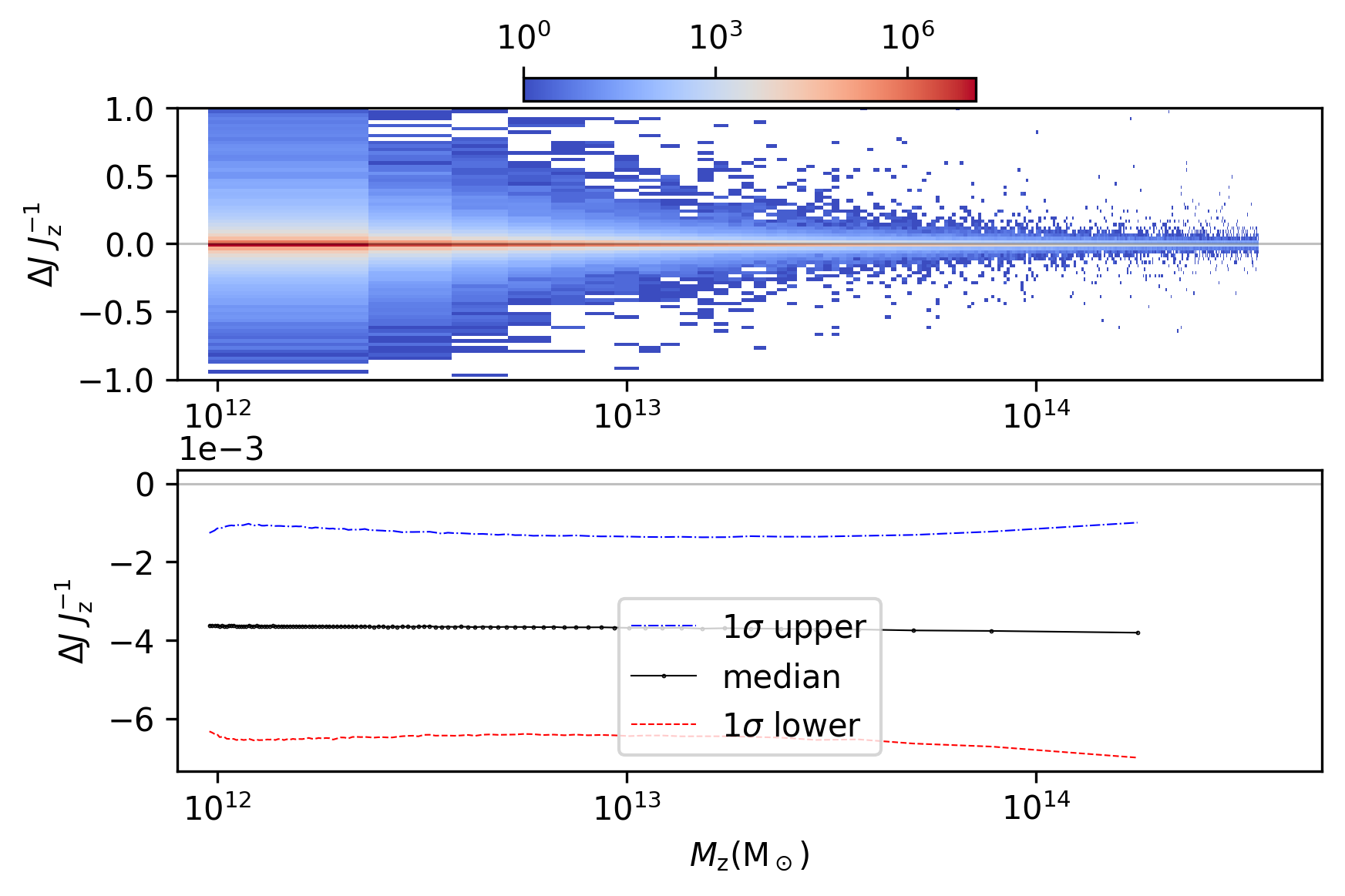}
	\caption{The $\Delta J/J_\mathrm{z}$ with respect to TianZero halo mass ($M_\mathrm{z}$). The total count is plotted in the upper panel with a color bar indicating the counting number. In the lower panel, the black dotted line denotes the medians of $\Delta J/J_\mathrm{z}$ among 100 mass bins, while the blue dashed-dotted and red dashed lines present the 1$\sigma$ region around the median. The upper panel uses the bins with the same width in the abscissa, and it is why the bin widths are visually smaller at higher mass. However, the lower panel uses the bins with equal halo amount, so the data points are more concentrated at lower mass. The identical discrepancies would arise in Figures \ref{fig4}, \ref{fig5} and \ref{fig13}.}\label{fig3}
\end{figure}

We plot the $\Delta J/J_\mathrm{z}$ ($\Delta J=J_\mathrm{\nu}-J_\mathrm{z}$, $J_\mathrm{\nu}$ is the halo modulus of TianNu and $J_\mathrm{z}$ is the halo modulus of TianZero) count in the upper panel of Figure \ref{fig3}, and calculate the 1$\sigma$ range enveloping the median among 100 mass bins with almost same volume in the lower panel of Figure \ref{fig3}.
$\Delta J/J_\mathrm{z}$ has the mean of -0.3717\% and the median of -0.3655\%, while the 89.71\% of them have smaller $J$ under the impact of neutrinos.

\begin{figure}[htb]
	\centering
	\includegraphics[scale=0.9]{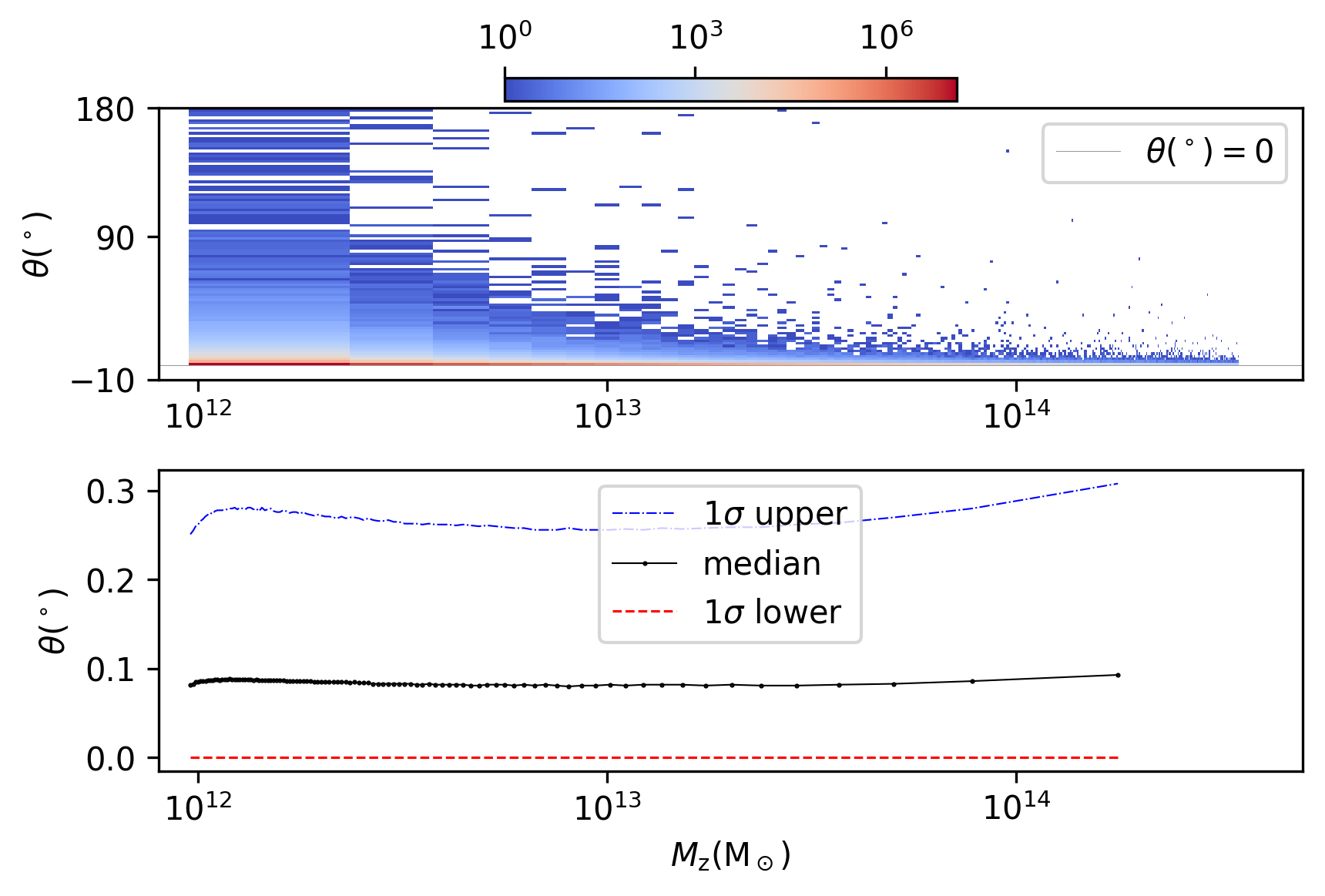}
	\caption{The $\Delta j/j_\mathrm{z}$ with respect to the $M_\mathrm{z}$. The total count is plotted in the upper panel with a color bar indicating the counting number. In the lower panel, the black dotted line denotes the median of $\Delta j/j_\mathrm{z}$ among 100 mass bins, while the blue dashed-dotted and red dashed lines present the 1$\sigma$ region around the median.}\label{fig4}
\end{figure}

Considering the different masses of the unit particle-groups in the two simulations, it is indirect to compare their AM modulus variation at a uniform mass and the traditional specific modulus is improper here. Instead, we re-scale their masses on a single particle-group and obtain the particle-mass-reduced (PMR) moduli, since they are the least simulated unit.
Therefore, we explore the PMR modulus, $\Delta j/j_\mathrm{z}$ ($\Delta j=j_\mathrm{\nu}-j_\mathrm{z}$, $j_\mathrm{z}=J_\mathrm{z}/N_\mathrm{z}$ and $j_\mathrm{\nu}=J_\mathrm{\nu}/N_\mathrm{\nu}$ , $N_\mathrm{z}$ and $N_\mathrm{\nu}$ are the amount of particle-groups forming a halo), and plot its distribution in Figure \ref{fig4}.
$\Delta j/j_\mathrm{z}$ has the mean of -0.5493\% and the median of -0.5650\%, while the 71.06\% of them have smaller $j$ under the impact of neutrinos.

\begin{figure}[htb]
	\centering
	\includegraphics[scale=0.9]{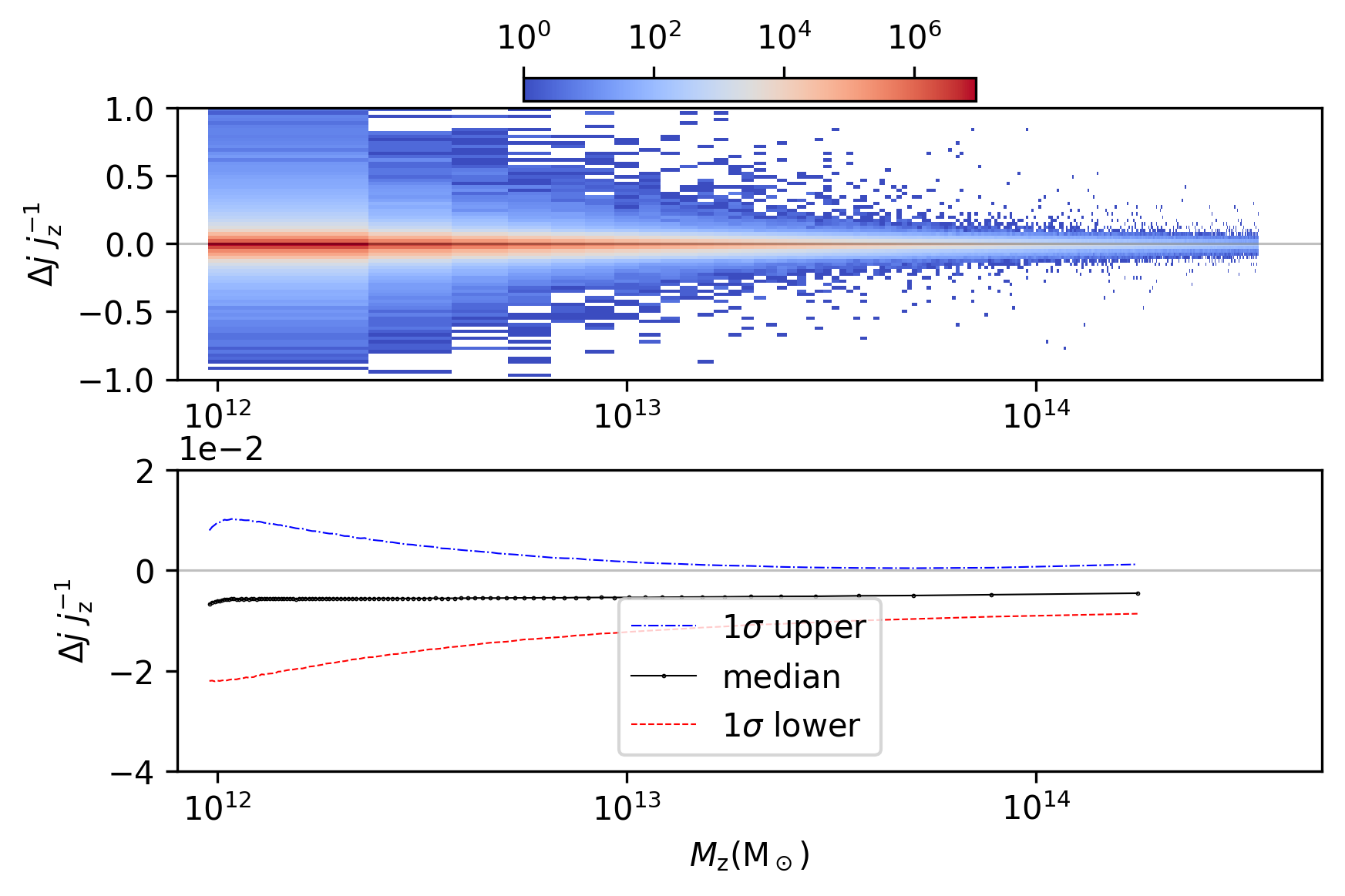}
	\caption{The orientation difference ($\theta$) with respect to $M_\mathrm{z}$. The total count is plotted in the upper panel with a color bar indicating the counting number. In the lower panel, the black dotted line denotes the median of $\theta$ among 100 mass bins, while the blue dashed-dotted and red dashed lines present the 1$\sigma$ region around the median.}\label{fig5}
\end{figure}

In Figure \ref{fig5}, we plot the directional angular difference $\theta$ (degree) between the $\bm{J_\mathrm{\nu}}$ and $\bm{J_\mathrm{z}}$:
\begin{equation}\label{eq2}
	\cos\theta=\frac{\bm{J_\mathrm{\nu}}\cdot\bm{J_\mathrm{z}}}{J_{\nu}J_z}=\frac{\sum_{i=1}^{3}J_{\mathrm{\nu},i} J_{\mathrm{z},i}}{J_\mathrm{\nu} J_\mathrm{z}}.
\end{equation}
They have a mean of 0.1646 degrees and a median of 0.0840 degrees, while 95.44\% of them are smaller than 0.6500 degrees under the impact of neutrinos.

\section{The Influence of Neutrino and Local Density}\label{sec3}
After investigating the relative modulus and direction variations from the TianNu and TianZero sets independently, we then consider the halos AM moduli with their local density environment. In the two simulations, we obtain the general halo number density of $1.406\times 10^{-2}$$h^3$Mpc$^{-3}$.

Given the irregular and fibrous condensed structures of dark halos, it is hard to determine their local density precisely by defining an inspected volume with consistent shape and magnitude. Additionally, the enormous amount of halos in data also reduces the efficiency of searching them by pairs.

To avoid a computing-expensive global searching among all dark halos, we first partition the tremendous whole box into many coarse cubes with the side length of $50h^{-1}\mathrm{Mpc}$. Considering that $100h^{-1}\mathrm{Mpc}$ ($10h^{-1}\mathrm{Mpc}$) is regarded as an upper (lower) boundary of smooth perturbed large-scale (small-scale) structures, the side length between the two values can be proper theoretically.
\sout{After division, we compute the matter density for each cube and plot the cube density distribution of TianZero in the old Figure 6.}

\begin{figure*}[htb]
	\centering
	\begin{minipage}{0.465\textwidth}
		\centering 
		\includegraphics[scale=0.9]{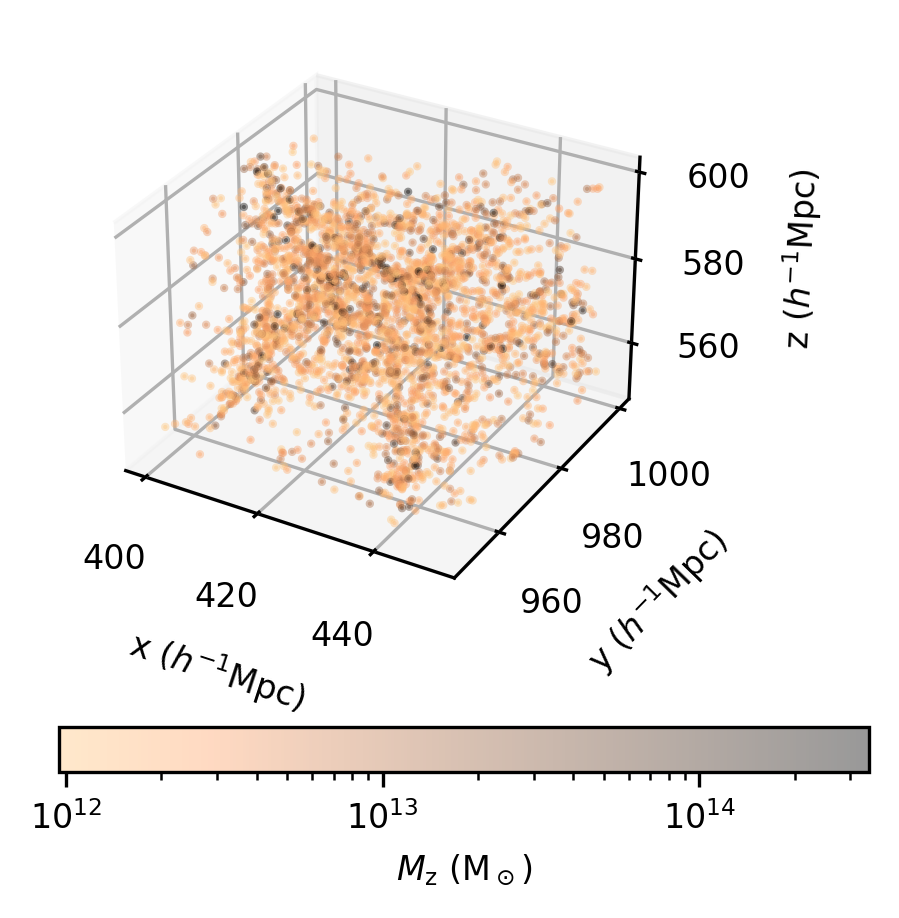}
		\caption{The densest cube in the TianZero box.}\label{fig7}
	\end{minipage}
	\begin{minipage}{0.05\textwidth}
	\end{minipage}
	\begin{minipage}{0.05\textwidth}
	\end{minipage}
	\begin{minipage}{0.05\textwidth}
	\end{minipage}
	\begin{minipage}{0.05\textwidth}
	\end{minipage}
	\begin{minipage}{0.465\textwidth}
		\centering 
		\includegraphics[scale=0.9]{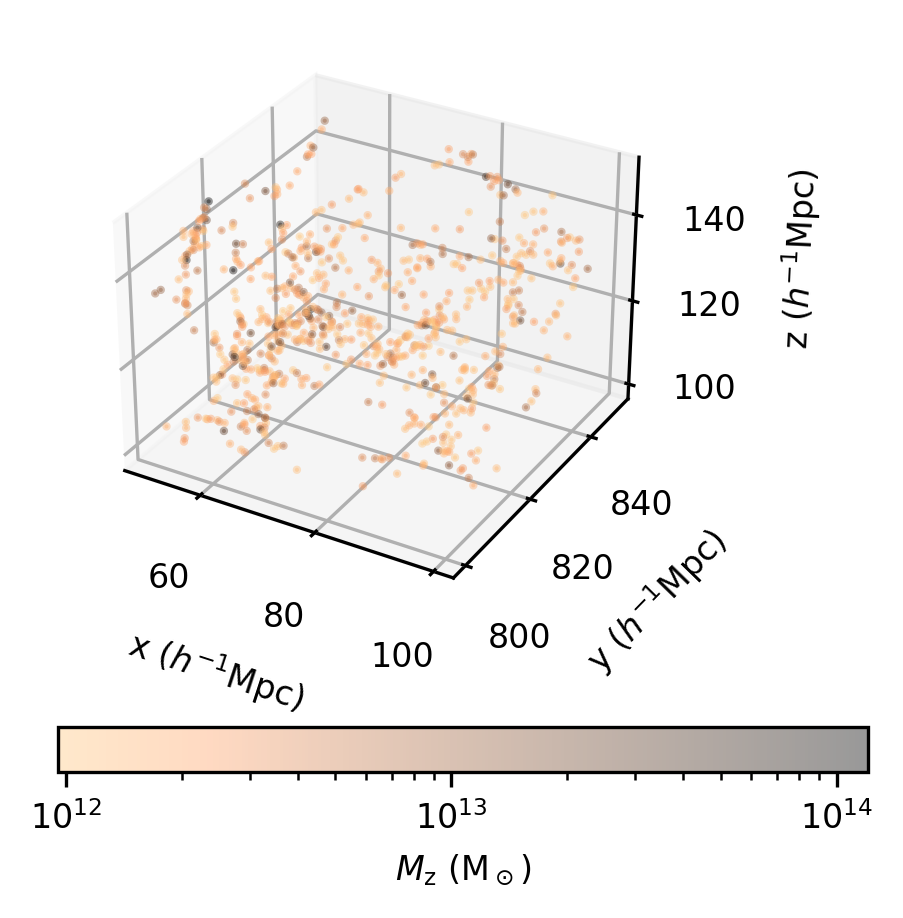}
		\caption{The sparsest cube in the TianZero box.}\label{fig8}
	\end{minipage}
\end{figure*}

We then select a cube to explore the halo AM modulus variation with different local densities and whether the neutrinos are added. For instance, in the selected cube with maximum (or minimum) matter density plotted in Figure \ref{fig7} (and Figure \ref{fig8}), we search the 360 halos located in the most compact as well as in the most scarce cells. Every cell is a sphere with a radius of $5h^{-1}\mathrm{Mpc}$ and centered in one halo, which is at least $5h^{-1}\mathrm{Mpc}$ away from the nearest block boundary. To avoid confusion in this paper, we emphasize that we use the box, cube and cell, to denote the space with $1200h^{-1}\mathrm{Mpc}$ side length, $50h^{-1}\mathrm{Mpc}$ side length and $5h^{-1}\mathrm{Mpc}$ radius respectively and strictly. For the same block in TianZero and TianNu, we obtain two sets of halos in compact and scarce local regions separately, and call them neoC (Compact-cell halo set in TianNu), neoS (Scarce-cell halo set in TianNu), zeoC (Compact set in TianZero) and zeoS respectively.
To collect the halos both in TianNu or TianZero, we select the first 300 halos matched from neoC-zeoC and neoS-zeoS, and name the four datasets as NeoC, NeoS, ZeoC and ZeoS.

To include more halos in more cubes for a statistical review, we select 300 compact and 300 scarce common halos from each of the 20 densest and 20 sparsest cubes respectively, summing up to four 6000-halo subsets representing the extended NeoC, NeoS, ZeoC and ZeoS sets.

\begin{figure*}[htb]
	\centering
	\includegraphics[scale=0.75]{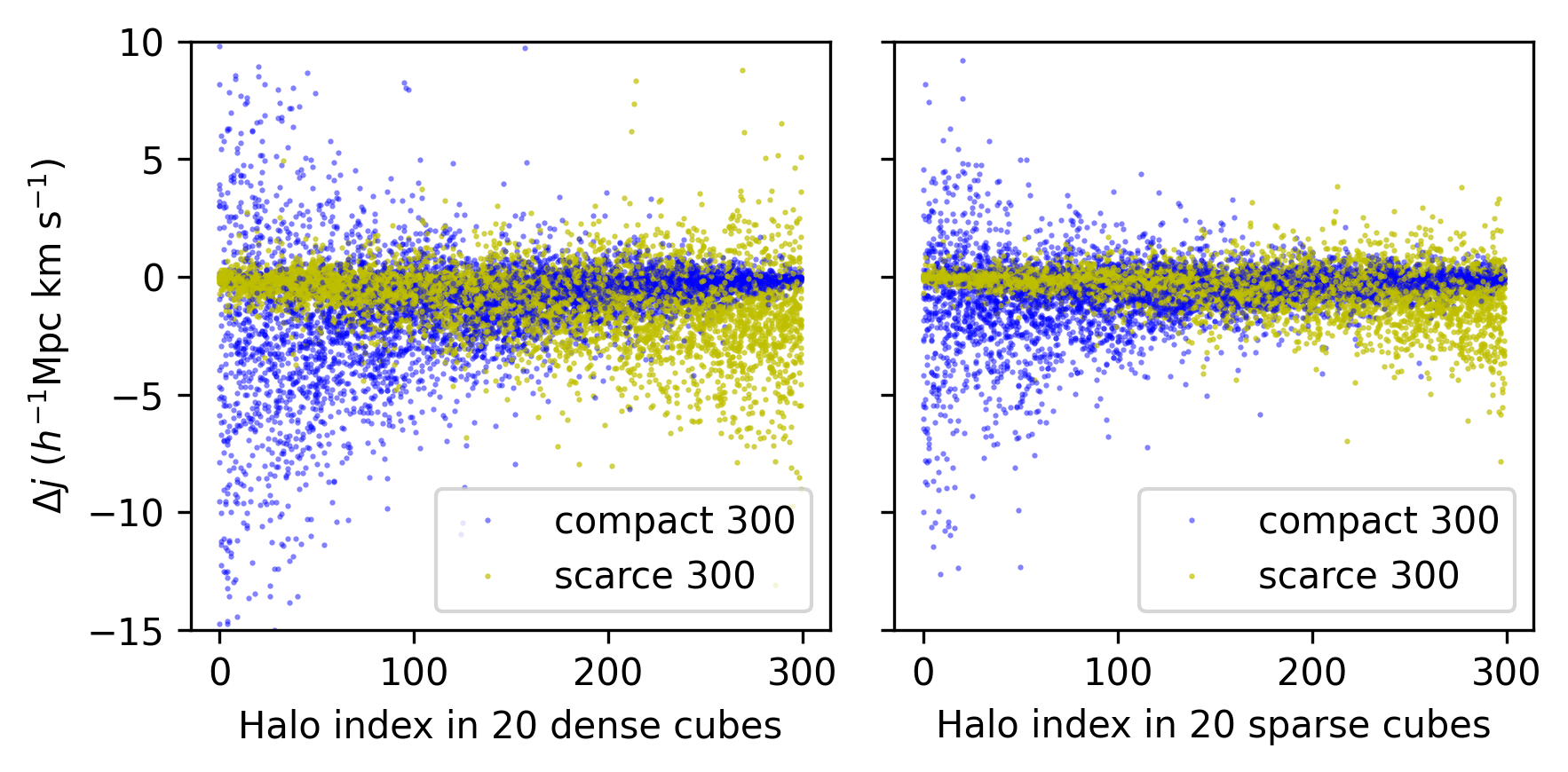}
	\caption{The variation of PMR moduli ($\Delta j$) of halos in extreme cubes. The left (or right) panel shows 6000 data points (300 halos from 20 sets) from the 20 densest (or sparsest) cubes. Less halo index indicates it has a more extreme (bigger or smaller) density environment.}\label{fig9}
\end{figure*}

\begin{figure*}[htb]
	\centering
	\includegraphics[scale=0.75]{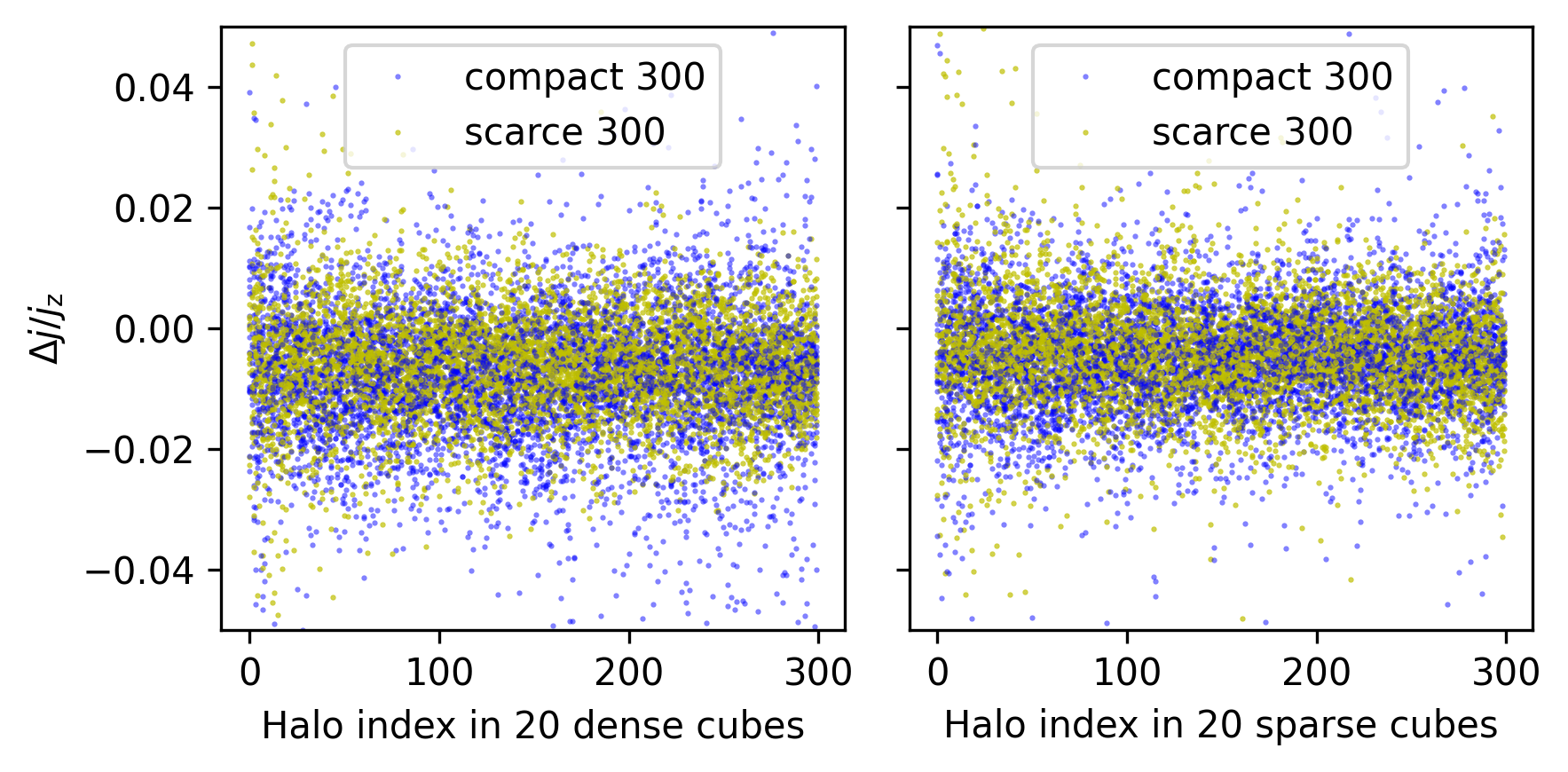}
	\caption{The relative variation of PMR moduli ($\Delta j/j_\mathrm{z}$) of halos in extreme cubes. The left (or right) panel shows 6000 data points (300 halos from 20 sets) from the 20 densest (or sparsest) cubes. Less halo index indicates it has a more extreme (bigger or smaller) density environment. }\label{fig10}
\end{figure*}

We compute each PMR modulus $j$ for each halo, and show the distribution of $\Delta j$ and $\Delta j/j_\mathrm{z}$ in the 20 densest and 20 sparsest cubes individually in Figures \ref{fig9} and \ref{fig10} separately. As for $\Delta j$, the halos in more compact cells (smaller index for compact halos and bigger index for scarce halos) tend to have more notable and negative $\Delta j$ values. However, the $\Delta j/j_\mathrm{z}$ seems more averagely negative for each particle-group.

\begin{figure*}[htb]
	\centering
	\includegraphics[scale=0.75]{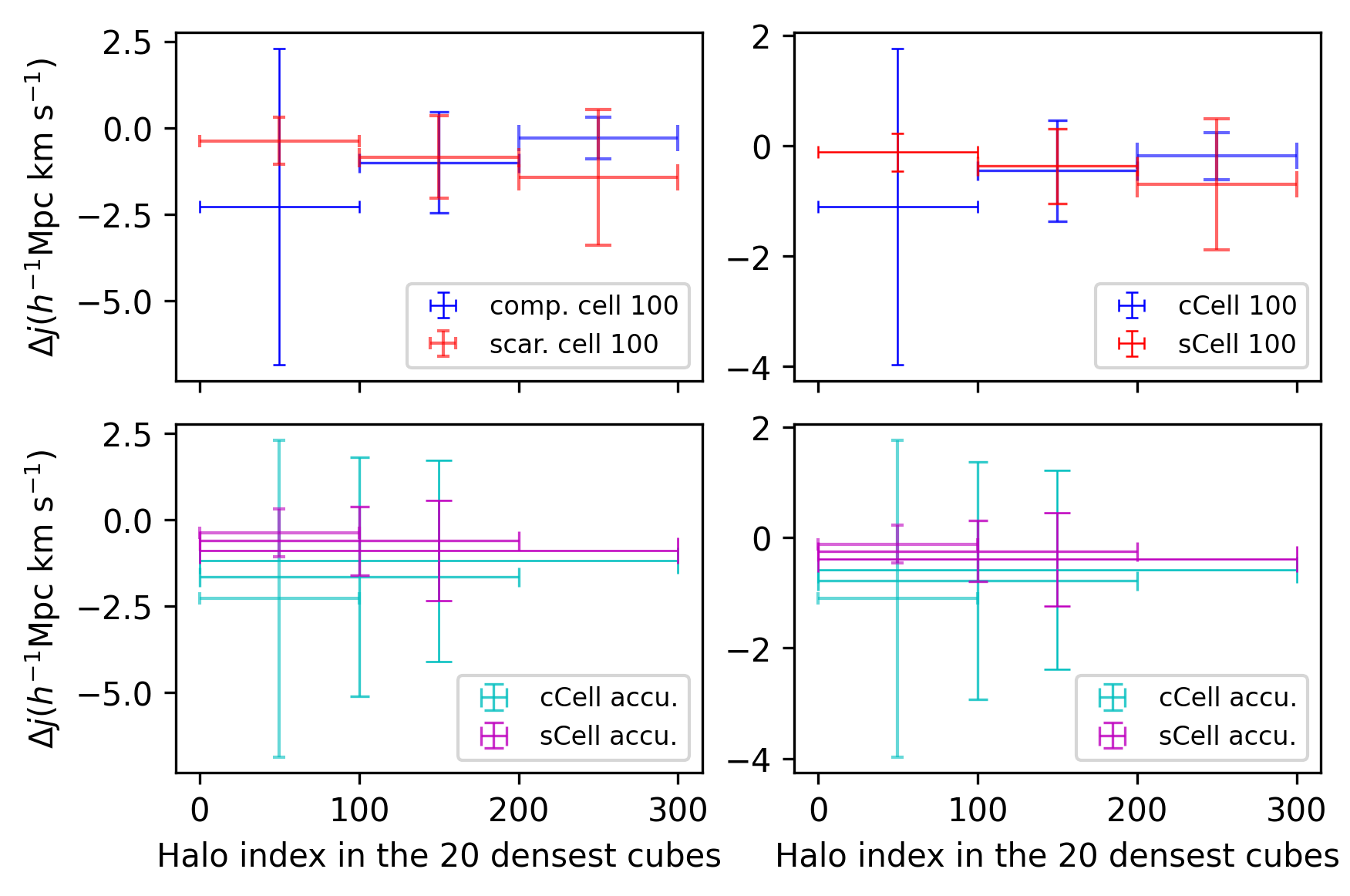}
	\caption{The statistics of $\Delta j$. The left (right) panels show the case of the densest (sparsest) cubes. The upper panels show the mean (center point) and standard deviation (vertical error bar) of every 100 halos, and the blue (red) signals denote the halos from compact (scarce) cells. The lower panels show the mean and standard deviation of the first 100, 200 and 300 halos, and the cyan (magenta) signals denote the halos from compact (scarce) cells. Less halo index indicates it has a more extreme (bigger or smaller) density environment. }\label{fig11}
\end{figure*}

\begin{figure*}[htb]
	\centering
	\includegraphics[scale=0.75]{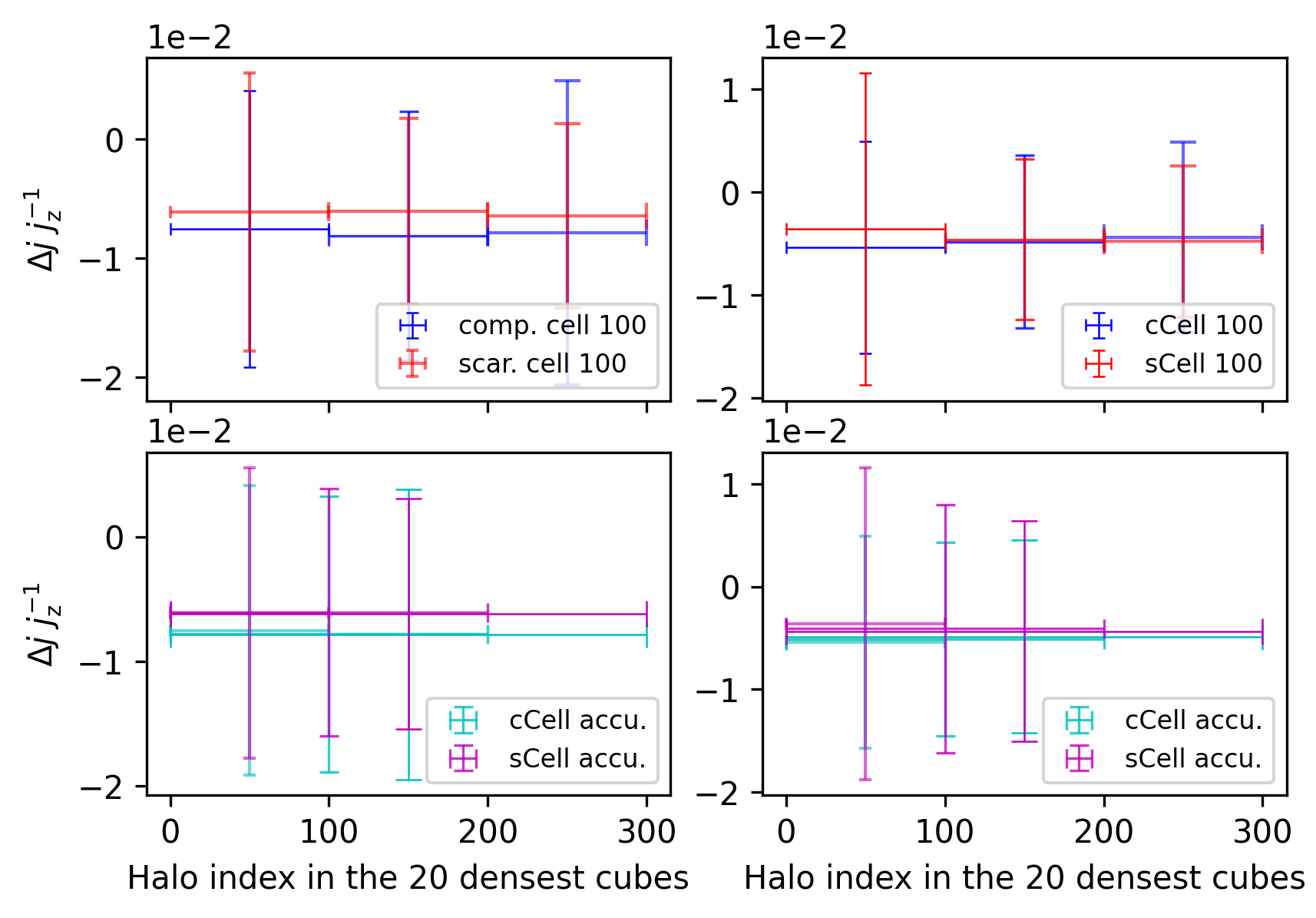}
	\caption{The statistics of $\Delta j$. The left (right) panels show the case of the densest (sparsest) cubes. The upper panels show the mean (center point) and standard deviation (vertical error bar) of every 100 halos, and the blue (red) signals denote the halos from compact (scarce) cells. The lower panels show the mean and standard deviation of the first 100, 200 and 300 halos, and the cyan (magenta) signals denote the halos from compact (scarce) cells. Less halo index indicates it has a more extreme (bigger or smaller) density environment. }\label{fig12}
\end{figure*}

Moreover, we show the mean and standard deviation of $\Delta j$ and $\Delta j/j_\mathrm{z}$ correspondingly in Figures \ref{fig11} and \ref{fig12} with bin-mode (upper panels) and accumulated-mode (lower panels).
The densest 100 compact and 100 scarce halos have similar $\Delta j$, although the former have doubled stand deviations.
The same situation occurs in the sparsest halos which have smaller stand deviations.
With more halos, extreme samples take up less proportion, and the two mean $\Delta j$ (for compact and scarce halos) approach closer to each other.
The $\Delta j/j_\mathrm{z}$ statistics are entirely indistinguishable in the two comparison modes.

\section{discussion}\label{sec4}
The distribution of halo's AM moduli ($J$) in Figure \ref{fig2} displays a plausible left-skewed lognormal distribution. It is obscure to explain why the $J$ distribution has a maximum value and this peak also arises from \cite{ueda94pasj}, \cite{catel96mn}, \cite{bosch02apj} and \cite{liao17apj}.

The relative halo moduli variation ($\Delta J/J_\mathrm{z}$, Figure \ref{fig3}) and AM orientation variation ($\theta$, Figure \ref{fig5}) are roughly the same for all-mass-range halos according to their 1$\sigma$ regions.
Through this indicator, only the relative PMR moduli variation ($\Delta j/j_\mathrm{z}$, Figure \ref{fig4}) is more obvious for low-mass halos, which have fewer CDM particle-groups and spatial volumes and are consequently more easily affected.
And it is noteworthy that a few outliers still occur with huge variations.

\begin{figure}[htb]
	\centering
	\includegraphics[scale=0.9]{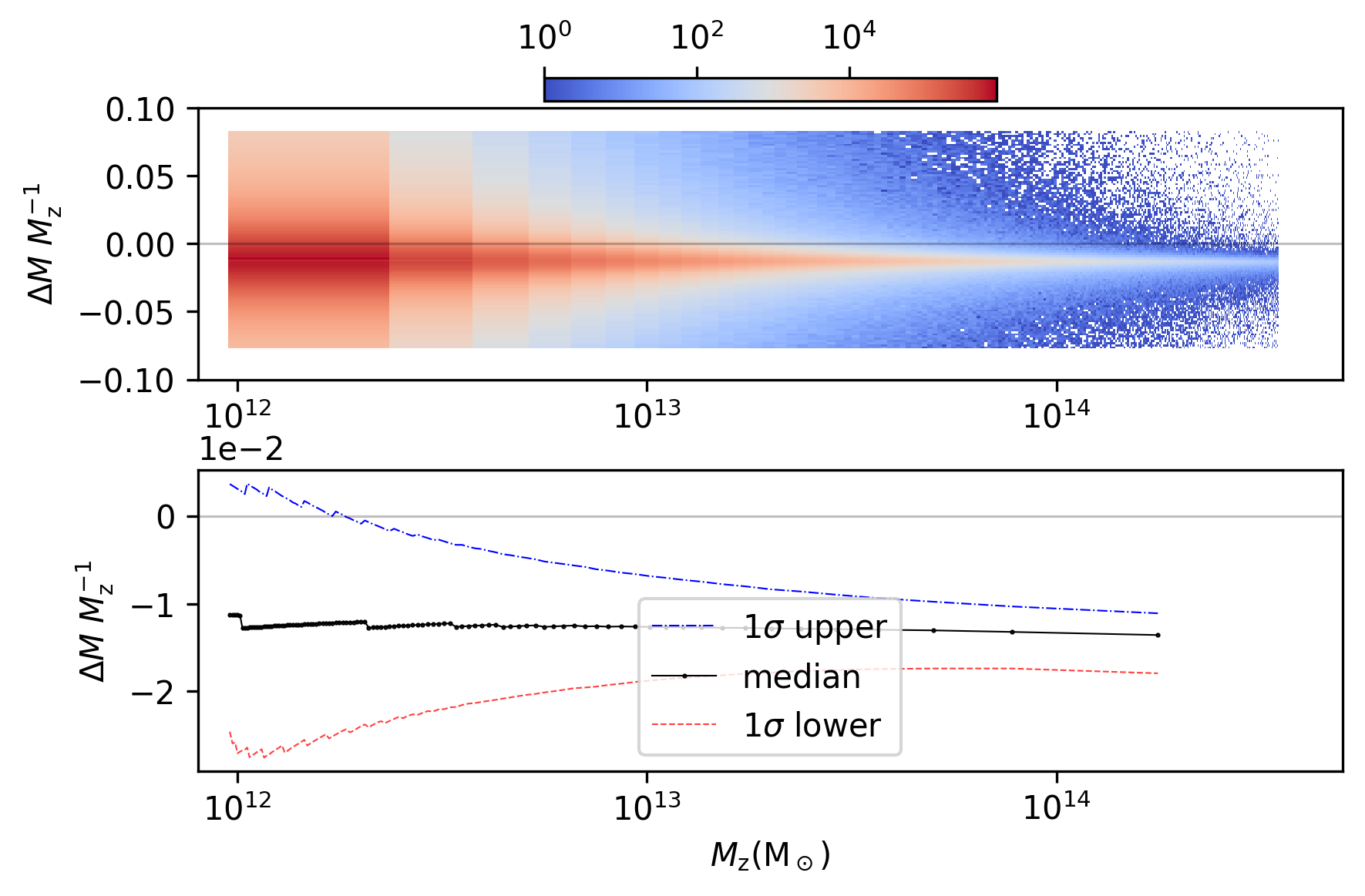}
	\caption{The $\Delta M/M_\mathrm{z}$ distribution. The ordinate is limited in [-0.08,0.08] because we set the 8\% boundary of mass variation in the halo-pair matching process. This truncation and discontinuous mass unit all contribute to the oscillation in our lower-mass statistic. The total count is plotted in the upper panel with a color bar indicating the counting number. In the lower panel, the black dotted line denotes the median of $\Delta M/M_\mathrm{z}$ among 100 mass bins, while the blue dashed-dotted and red dashed lines present the 1$\sigma$ region around the median.}\label{fig13}
\end{figure}

Over 70\% of $\Delta J/J_\mathrm{z}$ and $\Delta j/j_\mathrm{z}$ are less than 0, showing that the halo and PMR moduli of TianNu halos are smaller than the TianZero's in more cases.
However, the lower negative-proportion of $\Delta j/j_\mathrm{z}$ than the minus part of $\Delta J/J_\mathrm{z}$ indicates there should be some extra escape of particles-groups (and more mass loss) with massive neutrinos to decrease more halo AM modulus and make extra compensation to $\Delta J/J_\mathrm{z}$.
Therefore we plot the percent mass variation ($\Delta M/M_\mathrm{z}$) in Figure \ref{fig13}.
$\Delta M/M_\mathrm{z}$ has a mean of -1.2120\% and median of -1.2574\%, and 86.60\% of them is less than 0, supporting the escape.

There is an interesting phenomenon that $\Delta j$ involves halo's local density (Figure \ref{fig11}), while $\Delta j/j_\mathrm{z}$ more refers halo's local density (Figure \ref{fig4}) instead of its mass. The former can be explained because many low-mass halos counted in the cell of a massive central halo are affected by these gravitationally gathered massive neutrinos. And the latter reflects that the low-mass halo is a less-robust system since its relative PMR-modulus loss is more visible with the escape of every particle-group.

We also notice that \cite{liao17apj} carefully research the distribution of geometric parameters for AM modulus in a pure CDM particle simulation. However, we do not record the position, mass and AM of every inner particle. Therefore we cannot study like them in this paper.

The truncated $50h^{-1}\mathrm{Mpc}$ cubes are used to make mass evaluation faster and easier, taking the risk of ignoring some wall-near halos which may be located at a high-density cell of some massive central halo.
Besides, the most compact cells are usually centered in a massive CDM halo, while scarce cells usually include more low-mass halos.

The differential condensation effect \citep{yu17natast} exists between the scales of our finer cell ($10h^{-1}\mathrm{Mpc}$) and coarse cube ($50h^{-1}\mathrm{Mpc}$.), where neutrinos are more likely to cluster and collide at some dense regions, flushing away the gathered particles and diminishing the entire halo AM modulus.

TianNu was computed by CUBEP$^3$M in the Tianhe-2 supercomputer. Its halo algorithm has some shortcomings, such as the selection of the halo mass center, the computation of AM, the coarse halo-boundary identification in SO algorithm, the possible maximum gravitational error over 200\% \citep{harno13mn}, etc. These defects could be alleviated by newer and finer code, for example, the CUBE \citep{yu18apjs} which uses FoF halo finder and reduce the maximum gravitational error to 3.5\%. As a co-evolution of neutrino and CDM particles, few simulations, such as the cosmo-$\pi$ \citep{cheng20arx} and 4-trillion-grid Vlasov simulation \citep{yoshi21arx}, can compete with TianNu in the particle number, box size and its resolution simultaneously. It is hard to repeat the TianNu using the same or better conditions, and still needs more reliable data to reanalyze in further investigation.

\section{Conclusion}\label{sec5}
In this work, we study the dark halos' AMs, which are processed from TianNu CDM particle-neutrino co-evolution simulation with a flat $\Lambda$CDM model.

Investigating the halo modulus ($J$) , PMR modulus ($j$), mass and orientation variations due to neutrino-injection, we find the consistency that the 89.71\% of $\Delta J/J_\mathrm{z}$, 71.06\% of $\Delta j/j_\mathrm{z}$ and 86.60\% of $\Delta M/M_\mathrm{z}$ under neutrinos have negative values separately, indicating non-negligible systematic effects such as the free-streaming and differential condensation stimulated by massive neutrinos. Moreover, the 95.44\% orientation shifts are less than $0.65^{\circ}$.

Dividing the whole TianNu and TianZero box into coarse cubes, and searching the halos in finer local compact and scarce cells in the 20 densest and sparest cubes, we the distributions of $\Delta j$ and $\Delta j/j_\mathrm{z}$ for all selected 6000 halos.
In general, the $\Delta j$ varies and decreases more for the compact halos, and less for the scarce halos, indicating that $\Delta j$ relates to the halo's local density.
However, $\Delta j/j_\mathrm{z}$ more connects to halo's mass according to their 1$\sigma$ statistic instead of its local density, and more easily change in low-mass halos.

\normalem
\begin{acknowledgements}
We sincerely appreciate the editor and referee’s kind patience and so many suggestions, which help us greatly improve our manuscript. We especially thank Hao-Ran Yu for the discussion about the TianNu simulation and halo AM processing in CUBEP$^3$M. We also thank Cheng-zong Ruan, Kang Jiao and Jian Qin for their discussion and comments. This work was supported by the National Science Foundation of China (Grants No.11929301, 61802428) and National Key R\&D Program of China (2017YFA0402600).

The code used in this article is CUBEP$^3$M \citep{harno13mn,yu17natast}.

\end{acknowledgements}
  
\bibliographystyle{raa}
\bibliography{am_raa}

\end{document}